\documentstyle[preprint,prc,aps,floats,epsf]{revtex}

\tightenlines

\newcommand{\beq}{\begin{equation}}
\newcommand{\eeq}{\end{equation}}
\newcommand{\bea}{\begin{eqnarray}}
\newcommand{\eea}{\end{eqnarray}}
\newcommand{\ben}{\begin{eqnarray*}}
\newcommand{\een}{\end{eqnarray*}}

\def\kf{k_{\rm F}}
\def\kt{\widetilde k}
\def\pt{\widetilde p}
\def\qt{\widetilde q}
\def\dens{\rho}
\def\edens{{\cal E}}

\def\vec#1{{\bf #1}}

\def\nab{\overrightarrow{\nabla}}
\def\galnab{\tensor{\nabla}}

\begin{document}
\draft

\title{Effective Field Theory for Dilute Fermi Systems}

\author{H.-W.\ Hammer\thanks{{\tt hammer@mps.ohio-state.edu}}
        and R.~J.~Furnstahl\thanks{{\tt furnstahl.1@osu.edu}}}
\address{Department of Physics \\
         The Ohio State University,\ \ Columbus, OH\ \ 43210}

\date{April, 2000}

\maketitle

\begin{abstract}
The virtues of an effective field theory (EFT) 
approach to many-body problems are illustrated by
deriving the expansion for the energy of 
an homogeneous, interacting Fermi gas
at low density and zero temperature.
A renormalization scheme based on dimensional regularization with
minimal subtraction leads to a more transparent 
power-counting procedure and diagrammatic expansion
than conventional many-body approaches.
Coefficients of terms in the expansion with
logarithms of the Fermi momentum 
are determined by the renormalization
properties of the EFT that describes few-body scattering.
Lessons for an EFT treatment of nuclear matter are discussed.
\end{abstract}

\bigskip

\thispagestyle{empty}

\newpage

\section{Introduction}

The effective field theory (EFT) approach 
exploits the separation of scales in physical 
systems \cite{LEPAGE89,KAPLAN95,GEORGI94,EFT98,EFT99,Birareview}.
The application of EFT methods to many-body problems promises 
insight into the analytic structure of observables and a
consistent organization of many-body corrections, with reliable error
estimates.
Our ultimate goal is to calculate the properties of nuclear matter
and finite nuclei, 
but there are many issues as yet unresolved concerning the practical
application of EFT methods to these systems.
As a step toward this goal, 
we consider a simpler problem for which we can illustrate
some essential features that might be obscured in the full
nuclear matter treatment.

The dilute Fermi gas has often served as a benchmark many-body
problem, despite a shortage of physical applications.  
Recently, there has been experimental progress in creating and 
measuring dilute low-temperature Fermi systems \cite{DEMARCO99}, 
although these experiments at present are not sensitive to interactions.
In this paper, 
we consider a uniform system of nonrelativistic fermions of mass $M$,
interacting via a purely repulsive, spin-independent underlying interaction
with characteristic range $R \equiv 1/\Lambda$.
The restriction to repulsive interactions allows us to postpone discussion
of Cooper pair and liquid-gas instabilities \cite{PaB99,HAMMER00}.

The prototype interaction of this type is a hard-sphere repulsion of
radius $R$.
For momenta $p \ll \Lambda=1/R$, two-body scattering with a hard-sphere
potential is described by
the effective range expansion [see Eq.~(\ref{effrange})].
All of the effective range parameters are ``natural''; that is, their
magnitude is given by a power of the underlying scale $R$ 
times a dimensionless number of
order unity.
For general interactions,
natural values are expected for the effective range and 
higher-order parameters, but not necessarily for the scattering length,
which can be arbitrarily large if there is a bound state near zero energy.
The latter case is of particular interest, of course, because of the
nature of $s$-wave nucleon-nucleon scattering.
However, a fine-tuned scattering length requires special treatment
\cite{KSW98,UvK99},
so our discussion 
here will assume natural scattering lengths.

To our knowledge, the most complete low-density expansion
for the ground-state energy per particle
of such a system is \cite{AMUSIA68,BAKER71,BISHOP73}:
\bea
	\frac{E}{N} &=&
        \frac{\kf^2}{2M}
        \biggl[ \frac{3}{5} + (g-1)\biggl\{ \frac{2}{3\pi}(\kf a_s)
          + \frac{4}{35\pi^2}(11-2\ln 2)(\kf a_s)^2
          + \frac{1}{10\pi}(\kf r_s)(\kf a_s)^2
         \nonumber \\ & & \qquad \null
          + \bigl(0.076 + 0.057(g-3)\bigr)(\kf a_s)^3 \biggr\}
          + (g+1)\frac{1}{5\pi}(\kf a_p)^3
         \nonumber \\ & & \qquad \null
          + (g-1)(g-2)\frac{16}{27\pi^3}(4\pi-3\sqrt{3})(\kf a_s)^4
                  \ln(\kf a_s) + \cdots
        \biggr]  \ .
     \label{eq:EoverA}  
\eea      
In Eq.~(\ref{eq:EoverA}), $a_s$ and $r_s$ are the $s$--wave scattering
length and effective range, and $a_p$ is the $p$--wave scattering length.
The spin degeneracy is denoted by $g$.
For a natural system, this is an expansion in Fermi momentum
$\kf$ over the scale $\Lambda$.
The mean-field correction of ${\cal O}(\kf^3)$ dates 
from 1929 \cite{Len29},
the ${\cal O}(\kf^4)$ correction from the 1950's \cite{HuY57}, while the
${\cal O}(\kf^5)$ corrections and the logarithm were found in the 1960's
\cite{FETTER71}.
The complete expression in Eq.~(\ref{eq:EoverA}) has been derived using
the method of correlation functions \cite{AMUSIA68}, by expanding 
Goldstone diagrams \cite{BAKER71,BISHOP73}, and by expanding
Feynman diagrams \cite{BISHOP73}.
Here we rederive and illuminate this result using EFT methods.
   
While field theory approaches to many-body physics are well-established
 ({\it e.g.}, Refs.~\cite{FETTER71} and \cite{NEGELE88}),
 the treatment of divergences and renormalization issues
for a dilute Fermi system \cite{AMUSIA68,BAKER71,BISHOP73} are
reminiscent of the old-style approach to renormalization theory,
in which renormalization is primarily
a technical device to get rid of divergences in perturbation
theory ({\it i.e.}, to sweep problems under the rug)  \cite{BROWN93}.
The modern viewpoint in particle physics
that ``renormalization is an expression
of the variation of the structure of physical interactions with changes
in the scale of the phenomena being probed'' \cite{BROWN93} 
has led to powerful and elegant renormalization group methods and
to the development of the effective field theory approach,
which provides low-energy descriptions in terms of local, 
nonrenormalizable interactions.

We want to revisit the dilute Fermi gas with the same 
type of insight that effective field theory
has given to particle physics.
Some first steps were taken in Ref.~\cite{FST00}, in which a double
expansion in momentum and the strength of a model potential 
was exploited 
to study renormalization and EFT breakdown at finite density.
Here we extend the analysis to a general low-density expansion that is
perturbative only in the Fermi momentum.

Effective field theories 
for scattering of two fermions in the vacuum are now well understood
\cite{EFT98,EFT99,Birareview}
and are particularly straightforward for systems with natural scattering
lengths \cite{Kaplantalk,Biratalk}.
At low energies, with momentum $k \ll \Lambda$, the underlying
interactions
are not resolved, so they may be replaced by a series of local (contact)
interactions;
this description is completely general to the order of truncation.
Contributions to the scattering amplitude are expressed as a
systematic diagrammatic expansion in $k/\Lambda$.
The prescription for determining what diagrams 
contribute to a given order in the  expansion is called power counting.
Since the contact interactions are singular,
they require regularization
and renormalization, and the power counting prescription
depends on the subtraction scheme.
It is particularly useful to have a subtraction scheme in which
any Feynman graph with a particular vertex only contributes to
the expansion of the amplitude at a particular order, and 
for which power counting is simple dimensional analysis.
For a natural system, these conditions are satisfied by 
minimal subtraction implemented using
dimensional regularization \cite{Kaplantalk,Biratalk}.

The EFT expansion for two-body scattering efficiently reproduces
the effective range expansion,
which is a Taylor expansion in external
momentum $p$ divided by the 
characteristic scale of the underlying interaction $\Lambda$.  
When using a low resolution probe, only
the scale $\Lambda$ of new, unresolved physics enters.
Up to factors of order unity, the radius of convergence
of the expansion ({\it i.e.}, where errors are under control) is $\Lambda$, 
even though the domain of analyticity may be infinite.
A reliable estimate for the truncation error is order unity times 
$k/\Lambda$ to the first omitted power.

For a problem like this where the physics is ``transparent'',
the calculation would ideally be equally transparent.
The EFT treatment in the vacuum does exactly that. 
One might expect a simple analysis to be possible at finite
density as well, but the standard
treatments are actually quite complicated, with summations to all orders and
expansions for each diagram \cite{BAKER71,BISHOP73}.
The EFT approach presented here maintains the virtues of the EFT
description in the vacuum:  each Feynman diagram contributes to a definite
order in the $\kf/\Lambda$ expansion of the energy.
Previous work strained to eliminate divergences; with an EFT they are natural,
understood, and even exploited.

The analytic structure of the low-density expansion as a function
of $\kf$ is dictated by the interplay of short- and long-distance behavior.
This is particularly well manifested 
in the effective field theory approach.      
Recently, Braaten and Nieto\cite{BRAATEN97} showed
how renormalization in a dilute {\it boson\/} gas
determines the coefficients of logarithms in the low-density expansion of 
the energy.  Here we make an analogous analysis for fermion
systems to find the coefficient of the logarithm in Eq.~(\ref{eq:EoverA})
from renormalization group arguments.  
Extensions of this analysis may be fruitful in deriving
the analytic structure of the energy functional of
nuclear matter and other strongly interacting fermion systems.
We also discuss how to complete the renormalization, which {\it requires\/}
additional input beyond what is determined by two-body scattering.

In section II, we review how 
EFT power counting in the vacuum generates the effective range expansion.
In section III, the EFT is used to reproduce the analytic terms in
Eq.~(\ref{eq:EoverA}) and in section IV the logarithm is discussed.
We focus on the physics with a minimum of formalism.
More complete details will be given elsewhere, including numerical details
and extensions to finite temperature and other observables \cite{HAMMER00}.
The results are put in the context of the nuclear matter problem in Section~V.

\section{EFT at Zero Density}
\label{FSEFT}

In this section, we  review the EFT for two-body
scattering in the vacuum.%
\footnote{For a more complete treatment, 
see Refs.~\protect\cite{Kaplantalk,Biratalk}.} 
We consider
a natural EFT for heavy nonrelativistic fermions of mass $M$, whose 
interactions are spin-independent and
with strength correctly estimated 
by naive dimensional analysis based on an underlying scale $\Lambda$. 
In the center of momentum frame, the 
scattering amplitude $T$ of two particles with momenta of magnitude
$k$ and scattering angle $\theta$ can be expanded in partial waves as 
\beq
     T(k,\cos\theta) = \frac{4\pi}{M}\sum_l \frac{2l+1}{k \cot\delta_l-ik} 
                             P_l(\cos\theta)\ ,
   \label{tdef}
\eeq
where $\delta_l$ is the scattering phase shift for angular momentum $l$ and 
$P_l$ is a Legendre polynomial. Furthermore, for short-range
interactions, $k\cot\delta_l$ can be expanded in powers of $k^2$ as
\beq
\label{effrange}
k\cot\delta_0 = -\frac{1}{a_s}+\frac{1}{2} r_s k^2+\ldots\ ,\qquad
k\cot\delta_1 = -\frac{3}{k^2 a_p^3}+\ldots\ ,
\eeq
where $a_s$, $a_p$, and $r_s$ are the $s$- and $p$-wave scattering
lengths and $s$-wave effective range, respectively.
Inserting Eq.~(\ref{effrange}) into Eq.~(\ref{tdef}) and  expanding the
result up through ${\cal O}(k^2/\Lambda^2)$, we obtain
\beq
  T(k, \cos\theta) = 
         -\frac{4\pi a_s}{M}\left[ 1-i a_s k +(a_s r_s/2-a_s^2) k^2
      \right] -\frac{4\pi a_p^3}{M} k^2 \cos\theta + {\cal O}(k^3/\Lambda^3 )
	     \ .
         \label{texpand}
\eeq
For a natural system, $a_s \sim a_p \sim r_s \sim 1/\Lambda \equiv R$
and 
we have a {\em perturbative\/} expansion in $k/\Lambda$.
An example commonly studied is a hard-sphere gas with radius $R$,
in which case $a_s=a_p=R$ and $r_s=2R/3$.

Since we assume $k\ll\Lambda$, all interactions in the EFT are short-ranged
and we have only contact interactions.
Thus to reproduce this expansion with an EFT we consider
a general local Lagrangian for a nonrelativistic fermion
field that is invariant under Galilean, parity, and time-reversal 
transformations:%
\footnote{The extension to spin- and/or isospin-dependent 
interactions is straightforward and is illustrated in 
Ref.~\protect\cite{HAMMER00}.}
\bea
  {\cal L}  &=&
       \psi^\dagger \biggl[i\partial_t + \frac{\nab^{\,2}}{2M}\biggr]
                 \psi - \frac{C_0}{2}(\psi^\dagger \psi)^2
            + \frac{C_2}{16}\Bigl[ (\psi\psi)^\dagger 
                                  (\psi\galnab^2\psi)+\mbox{ h.c.} 
                             \Bigr]   
  \nonumber \\[5pt]
   & & \null +
         \frac{C_2'}{8} (\psi \galnab \psi)^\dagger \cdot
             (\psi\galnab \psi) 
         - \frac{D_0}{6}(\psi^\dagger \psi)^3 +  \ldots\,,
  \label{lag}
\eea
where $\galnab=\overleftarrow{\nabla}-\nab$ is the Galilean invariant
derivative and h.c.\ denotes the hermitian conjugate. 
The terms proportional to $C_2$ and $C_2'$ contribute to $s$-wave and
$p$-wave scattering, respectively.
The dots represent terms with more derivatives and/or more
fields.
Higher-order terms with time derivatives are omitted, as they can be
eliminated in favor of terms with spatial derivatives 
\cite{Kaplantalk,Biratalk}. 
Relativistic kinematic corrections can be included systematically, but
all such terms are
suppressed by factors of $(\Lambda/M)^2$, which we take to be negligible;
they are considered explicitly in Ref.~\cite{HAMMER00}.

\begin{figure}[t]
\begin{center}
  \epsfxsize=14.cm
  \epsffile{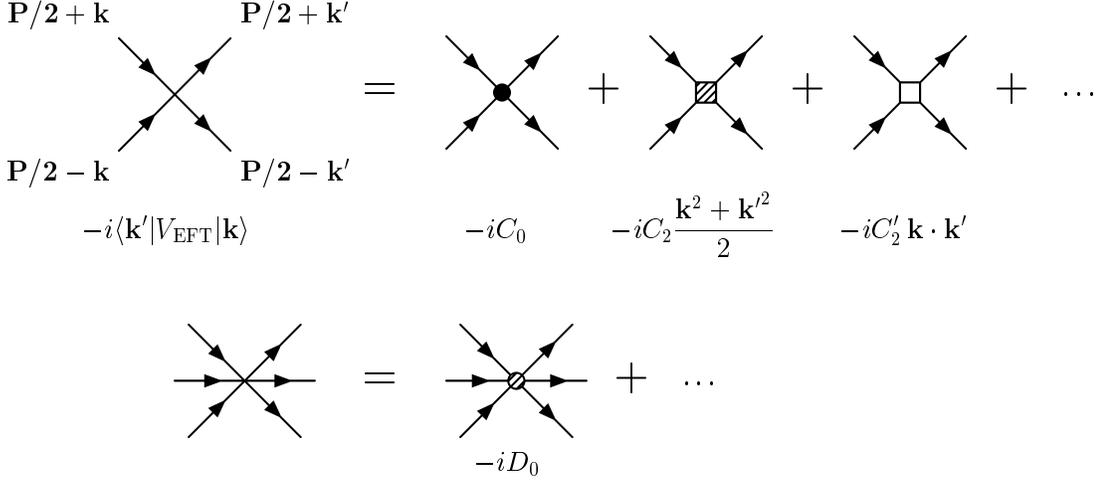}
\end{center}
\vspace*{-.2in}
\caption{Feynman rules for $\langle\vec{k}'|V_{\rm EFT}|\vec{k}\rangle$
and the leading 3-body contact interaction.
The spin indices have been suppressed.}
\label{eftvertex}
\end{figure}

The scattering amplitude for fermions in the vacuum is simply related
to a sum of Feynman graphs computed according to this Lagrangian.
The terms in Eq.~(\ref{lag}) 
involving only four fermion fields reduce in momentum 
space to simple polynomial vertices
that are equivalent to a momentum expansion of an EFT potential for 
particle-particle scattering,
\beq
  \langle \vec{k'} | V_{EFT} | \vec{k}\rangle
      = C_0 + C_2 (\vec{k'}^2 +\vec{k}^2)/2
         + C_2' \,\vec{k'}\cdot\vec{k} 
         + \ldots \ ,
  \label{effpot}
\eeq
(see Fig.~\ref{eftvertex} for the Feynman rules).
Because of Galilean invariance, the interaction depends only on the 
relative momenta  $\vec{k}$ and $\vec{k'}$ of the incoming and outgoing 
particles. 
The coefficients $C_0$, $C_2$, and $C_2'$ can 
be obtained from matching the EFT
to a more fundamental theory or to 
(at least) three independent pieces of experimental data.
It is important to note that Eq.~(\ref{effpot}) is {\em not\/}
simply the term-by-term momentum-space expansion of an underlying potential
because the coefficients also contain short-distance contributions
from loop graphs
(see Ref.~\cite{FST00} for a simple example of the matching and
renormalization of the coefficients).

The loop integrals that appear in Feynman graphs for two-body
scattering  with this EFT in $D$ spatial dimensions reduce to 
\beq
   L_n \equiv \int\! \frac{d^{D}q}{(2\pi)^{D}}\,
           \frac{q^{2n}}{k^2-q^2 + i\epsilon}  \ .
	\label{Ln}
\eeq
For $D=3$ spatial dimensions, these integrals are divergent and must be
regulated.  If we apply a sharp momentum cutoff $\Lambda_c$ on $q$,
evaluate the integral, and expand in $k$, we find the leading terms have
positive powers of $\Lambda_c$.  These are called power divergences.
For example, when $n=0$,
\beq
  L_0 \stackrel{D= 3}{\longrightarrow} -\frac{1}{2\pi^2}\Lambda_c
      - \frac{i}{4\pi} k + {\cal O}(k^2/\Lambda_c) \ .
	  \label{L0}
\eeq
The power divergences come from the short-distance 
region of the integral where the
loop momentum is of order $\Lambda_c$.  
This contribution is incorrect and depends on the details of how
we regulate the integral, but can be removed systematically
using local counterterms.
In contrast, contributions from loop momenta of order of the small
external momentum $k$ [the second term in Eq.~(\ref{L0})] are physical.
These contributions are completely decoupled; the $\Lambda_c$
coefficient tells us nothing about the $k$ coefficient.

If we imagine two-body scattering in $D=2$ dimensions,
the situation is different, because there will always be a logarithmic
divergence (as well as power divergences in general).
For example, when $n=0$, 
\beq
  L_0 \stackrel{D= 2}{\longrightarrow} \frac{1}{2\pi}\ln(k/\Lambda_c)
      + {\cal O}(1) \ .
	  \label{L20}
\eeq
The dependence on $\ln(\Lambda_c)$ is canceled by a counterterm,
but now the contributions from the different momentum regions are
coupled.  Since the logarithms of $k$ and $\Lambda_c$ must match,
the coefficient of the logarithm is independent of the regularization.
Further, it can be determined by looking only at the log divergence in the
integral, without a full calculation.  We will take advantage
of this simplification later when we consider three-body scattering,
in which logarithmic divergences appear for $D=3$. 

Since the power divergences do not provide useful information here
and complicate the power counting, it is appropriate to
use a
minimal subtraction scheme, which by construction subtracts the
power divergences.
Dimensional regularization with minimal subtraction is particularly
convenient for a natural EFT, 
since loop integrals for two-body scattering pick up {\em only\/} the
scale of the external momentum
(logarithms are also possible when more than
two fermions scatter).
In particular,\cite{Kaplantalk,Biratalk}
\beq
  L_n \stackrel{D\rightarrow 3}{\longrightarrow}
      - \frac{i}{4\pi} k^{2n+1} \ .
	  \label{LnDR}
\eeq
Thus, the underlying physical scale $\Lambda$ appears only in the
coefficients and the loops contribute only factors of $k$, which can
be determined by dimensional analysis.

Therefore 
the estimated
contribution of a given diagram is found by counting (i) for every 
propagator a factor $M/k^2$, (ii) for every loop integration
a factor of $k^5/M$, and (iii) for every $n$-body vertex with $2i$ 
derivatives a factor $k^{2i}/M \Lambda^{2i+3n-5}$.
This scaling of the coefficients with $M$
follows from Galilean invariance and with $\Lambda$ from
dimensional analysis \cite{Kaplantalk,Biratalk}. In particular, 
the vertices with four $(n=2)$ and six $(n=3)$ fields scale as
\beq
\label{scaling}
C_{2i} \sim {4\pi \over M \Lambda^{2i+1}}
\quad \mbox{and}\quad
D_{2i} \sim {4\pi \over M \Lambda^{2i+4}} \ ,
\eeq
which means that more complicated graphs are suppressed at
low energies by additional powers of $k/\Lambda$.
As above, $\Lambda$ is an ultraviolet scale which is determined by 
the physics not explicitly included in the EFT
(such as the mass of a heavy exchanged particle). 
For momenta $k\sim\Lambda$, 
the expansion in powers of $k/\Lambda$ breaks down and the new physics at 
the scale $\Lambda$ has to be explicitly included in the EFT.
Therefore, $\Lambda$ is also called the breakdown scale of the EFT.

We can use the topological properties of a graph to determine 
that a diagram with $L$ loops or $E$ external lines and
$V_{2i}^n$ $n$-body vertices with $2i$ derivatives scales precisely
as $k^\nu$, where 
\bea
\nu &=& 3L+2+\sum_{n=2}^\infty \sum_{i=0}^\infty (2i-2) V_{2i}^n
   \nonumber \\
   &=& 5 - \frac{3}{2}E + 
        \sum_{n=2}^\infty \sum_{i=0}^\infty (2i+3n-5) V_{2i}^n \ .
        \label{kpower}
\eea
At every order in $k/\Lambda$ there are a finite number of graphs
that contribute. The mass $M$ enters all Feynman diagrams in the same 
way as a factor $1/M$ and is irrelevant for the relative scaling of
diagrams.

\begin{figure}[t]
\begin{center}
  \epsfxsize=14.cm
  \epsffile{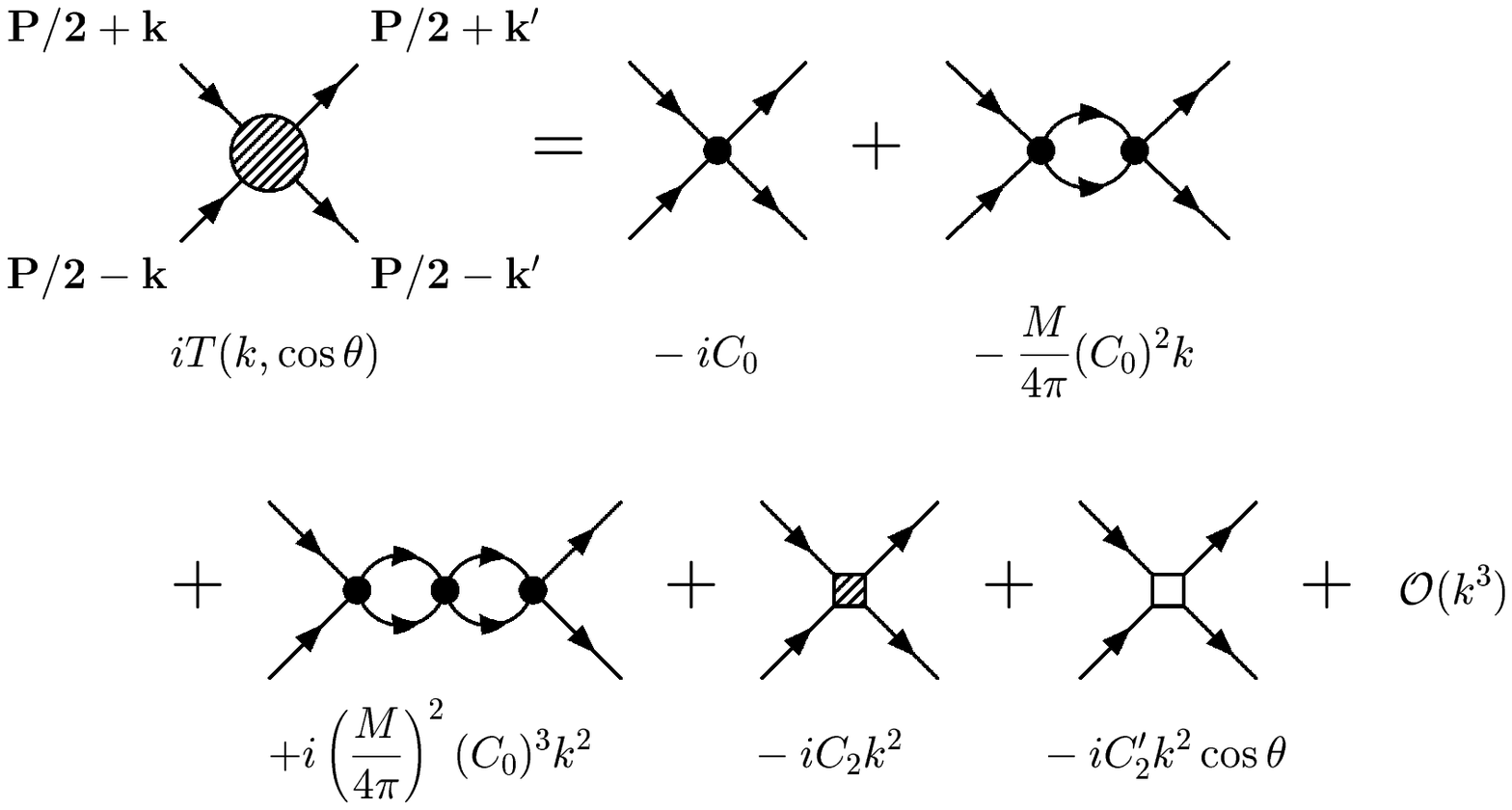}
\end{center}
\vspace*{-.2in}
\caption{Expansion of $T(k, \cos\theta)$ in the EFT.}
\label{eftere}
\end{figure}

Now we can determine the coefficients $C_{2i}$ by matching.
In our case, the experimental data we match to are the leading coefficients
of 
the effective range expansion for the scattering amplitude $T$,
Eq.~(\ref{texpand}).
Evaluating $T$ in the center-of-momentum frame of the two particles, 
we obtain (see Fig.~\ref{eftere}) 
\beq
\label{tcalc}
T(k, \cos\theta)=-C_0 \Bigl[ 1-i \frac{M}{4\pi}C_0 k + \frac{C_2}{C_0} k^2
-\left( \frac{M}{4\pi}C_0\right)^2 k^2  + \frac{C_2'}{C_0} k^2 \cos\theta 
\Bigr]+ {\cal O}(k^3/\Lambda^3 )\,,
\eeq
where $|\vec{k}|=|\vec{k'}|=k$ and $\theta$ is the scattering angle.
Matching Eq.~(\ref{tcalc}) to Eq.~(\ref{texpand}),
we can express the $C_{2i}$ in terms of the effective range parameters:
\beq
      C_0 = \frac{4\pi a_s}{M},\qquad C_2=C_0 \frac{a_s r_s}{2}, 
	       \quad \mbox{and}\quad
          C_2' = \frac{4\pi a_p^3}{M}
      \label{C2imatch}    \ .
\eeq
Thus the EFT reproduces the effective range
expansion in the vacuum.%
\footnote{If the scattering length is unnatural,
the power counting has to be
modified and the perturbation series must be resummed to avoid
a premature breakdown of the EFT. 
This case has received 
considerable attention recently \cite{EFT98,EFT99,Birareview}.}
With the interaction parameters in the 
Lagrangian expressed in terms of the effective range parameters, 
we are now ready to apply the EFT to dilute Fermi systems at finite
density.

\section{Dilute Fermi Systems}

We first review the
noninteracting Fermi gas at zero temperature to set our notation
and conventions. 
In this case,
all states up to the Fermi momentum $\kf$ are occupied while the
ones above $\kf$ are empty.
The density $\dens=N/V$ of a noninteracting
system with Fermi momentum $\kf$
and degeneracy $g=(2s+1)$ is obtained by summing over all occupied states,
\beq
\label{fermid}
\dens=g \int\! \frac{d^3 k}{(2\pi)^3} \theta(\kf-k)
      = \frac{g\,\kf^3}{6\pi^2}\ .
\eeq
In a uniform system, this relationship between $\dens$ and $\kf$ 
is unchanged by interactions \cite{FETTER71}.
The energy density $\edens_0$ of a noninteracting Fermi gas is given 
by weighting the integral in Eq.~(\ref{fermid}) 
with the free single particle energy
$\omega_{\vec{k}}\equiv\vec{k}^{\,2}/(2M)$, which leads to
\beq
\label{eden0}
       \edens_0 = \dens\,\frac{3}{5}\,\frac{\kf^2}{2M}\ .
\eeq
The energy per particle $E_0/N$ is simply $\edens_0/\rho$.

Equation~(\ref{eden0}) receives corrections from the interaction between the 
fermions. For low-density systems with short-range interactions,
$\dens\propto \kf^3 \ll 1/R^3 \sim \Lambda^3$, and  the bulk properties
of the system are amenable to an expansion in $\kf R$ or $\kf/\Lambda$.
As discussed earlier, the EFT method is ideally suited to deal
with such a separation of scales.
In this section, we present an EFT for low-density Fermi systems
and illustrate its application by calculating the corrections
to Eq.~(\ref{eden0}) to order $(\kf R)^3$.
The use of contact interactions and EFT power counting
makes the EFT treatment straightforward and 
more transparent than  traditional approaches ({\it e.g.}, see
Ref.~\cite{AMUSIA68}).

\begin{figure}[t]
\begin{center}
  \epsfxsize=13.cm
  \epsffile{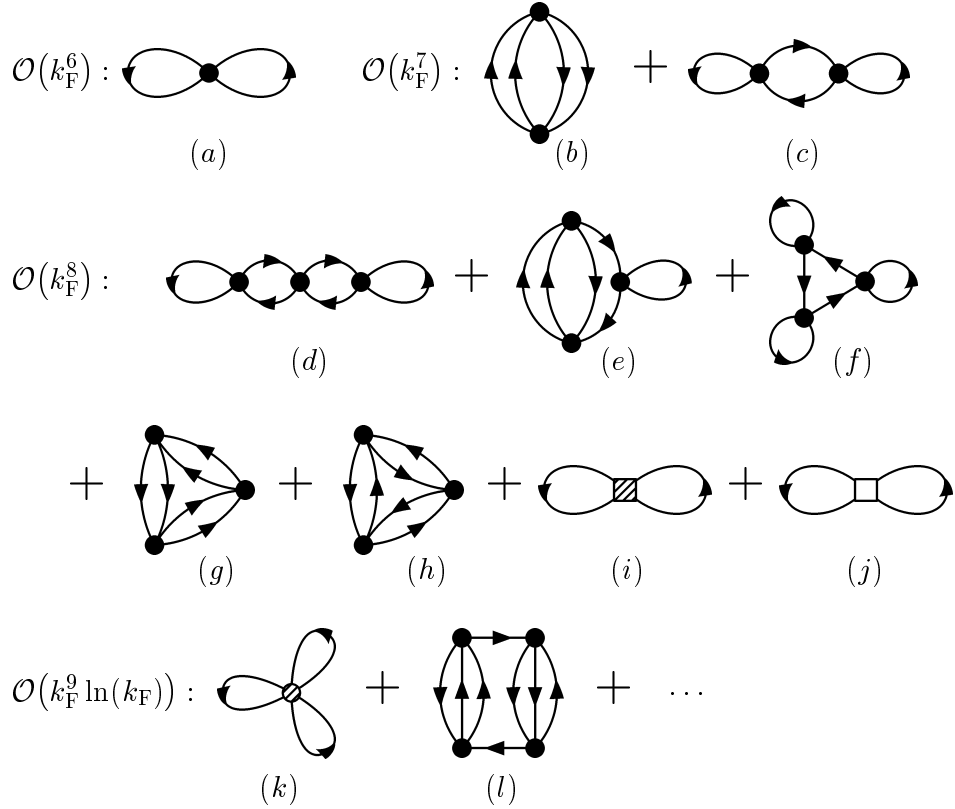}
\end{center}
\caption{Hugenholtz diagrams for the energy density ${\cal E}$
in a homogeneous, dilute Fermi system.  The vertices are defined
in Fig.~\protect\ref{eftvertex}.}
\label{fig_hugenholtz}
\end{figure}

The ground-state energy density ${\cal E}$ 
can be expressed as a sum of Hugenholtz diagrams,
which are  closed, connected
Feynman diagrams with appropriate symmetry factors
(see Fig.~\ref{fig_hugenholtz}).
To apply the EFT from the previous section at finite density, 
we have to adapt the
power-counting rules from the vacuum to the medium.
This is immediate with dimensional regularization and minimal subtraction,
because the
typical momentum flowing in {\it both\/} particle and hole lines in energy
diagrams is of order $\kf$ (see below).
The result is clear for hole lines, but particle lines 
can in principle have momenta that are unbounded  
or as large as an imposed cutoff $\Lambda_c$.
However, as discussed above,
the contributions of states with momenta close to the cutoff,
which depends on the regularization scheme in an EFT, are removed
by minimal subtraction.
The correct
effect of high-momentum modes on low-energy physics is incorporated in the 
coefficients of the local operators in the Lagrangian instead.
The end result is that 
we can use Feynman diagrams for the ground-state energy, 
in which holes and particles are propagated together, and apply the 
vacuum power counting with $\kf$ playing the role of the external momentum. 

Thus we estimate the contribution of a given 
Hugenholtz diagram by counting (i) for every 
propagator a factor $M/\kf^2$, (ii) for every loop integration
a factor of $\kf^5/M$, and (iii) for every $n$-body vertex with $2i$ 
derivatives a factor $\kf^{2i}/M\Lambda^{2i+3n-5}$.
Again, the mass $M$ enters all Feynman diagrams in the same way as
a factor $1/M$.
Furthermore, the range $R$ always has to match up with $\kf$ to produce 
the dimensionless expansion parameter $\kf R$. The correct power of $R$ 
can be inferred from the overall dimension of the diagram, which
is an energy density. Consequently, it is sufficient to keep
track of the powers of $\kf$ to classify the relative size
of diagrams. 
The  power counting rules are conveniently
summarized in one equation in terms of the vertices of a given diagram.
In particular, a graph with $V_{2i}^n$ $n$-body vertices 
with $2i$ derivatives scales as $\kf^\nu$, where $\nu$ is given by
\beq
   \nu = 5  + 
        \sum_{n=2}^\infty \sum_{i=0}^\infty (2i+3n-5) V_{2i}^n
        \label{kfpower}
\eeq
[this follows from Eq.~(\ref{kpower}) with $E=0$].
As at zero density, diagrams with more vertices and more derivatives  are 
suppressed, and therefore the EFT expansion is efficient and systematic.

The leading order Hugenholtz diagrams contributing 
to the energy density are shown in Fig.~\ref{fig_hugenholtz}.
From Eq.~(\ref{eden0}),
the energy density of the noninteracting Fermi gas is of
${\cal O}(\kf^5)$; the first correction Fig.~\ref{fig_hugenholtz}(a) 
is of ${\cal O}(\kf^6)$. 
The Feynman rules for the contributions to the
energy density from the diagrams in Fig.~\ref{fig_hugenholtz}
are essentially the same as in standard many-body treatments
({\it e.g.}, see Ref.~\cite{FETTER71,NEGELE88}),
which we summarize here.

First, draw all distinct, fully connected diagrams contributing to 
a given order in $\kf$ [as determined by Eq.~(\ref{kfpower})].
Distinct diagrams are those that cannot be deformed to coincide
with each other, including the direction of arrows.
To evaluate a diagram:
\begin{enumerate}
\item Assign nonrelativistic four-momenta (frequency and three-momentum) 
to all lines and enforce four-momentum conservation
at each vertex.
\item 
For each vertex include a factor $-i$ times the corresponding term
from the effective potential $\langle\vec{k}'|V_{\rm EFT}|\vec{k}\rangle$, 
as shown in Fig.~\ref{eftvertex}.
(Recall that ${\bf k}$ and ${\bf k'}$ are {\em relative\/} momenta.) 
The spin structure of the vertices 
has been suppressed in Fig.~\ref{eftvertex}.
For spin-independent interactions, the two-body vertices have
the structure $(\delta_{\alpha\gamma}\delta_{\beta\delta}
+\delta_{\alpha\delta}\delta_{\beta\gamma})$,
where $\alpha,\beta$ are the spin indices of the incoming lines
and $\gamma,\delta$ are the spin indices of the outgoing lines.
For spin-dependent interactions one inserts the appropriate operator
in spin space at each vertex instead.

For each internal line include a factor 
$iG_0 (\kt)_{\alpha\gamma}$, where
$\kt \equiv (k_0,\vec{k})$ is the four-momentum assigned to the line,
$\alpha$ and $\gamma$ are spin indices, and
\beq
    iG_0 (\kt)_{\alpha\gamma}=i\delta_{\alpha\gamma}
    \left( \frac{\theta(k-\kf)}{k_0-\omega_{\vec{k}}+i\epsilon}
      +\frac{\theta(\kf-k)}{k_0-\omega_{\vec{k}}-i\epsilon}\right) \ .
      \label{freeprop}
\eeq

\item Perform the spin summations in the diagram. 
In every closed fermion loop, substitute $-g$ for each
$\delta_{\alpha\alpha}$.

\item Integrate over all independent momenta with a factor 
$\int\! d^4 k /(2\pi)^4$
where $d^4k\equiv dk_0\,d^3k$. 
If the spatial integrals are divergent, they are defined in $D$ spatial
dimensions and renormalized using minimal subtraction as discussed in
Sect.~\ref{FSEFT} and Ref.~\cite{Kaplantalk,Biratalk}.
For lines ending and originating at the same
vertex, multiply by $\exp(ik_0\eta)$ and take the limit $\eta\to 0^+$
after the contour integrals have been carried out. This procedure
automatically takes into account that such lines must be hole lines.

\item Multiply by a symmetry factor $i/(S \prod_{l=2}^{l_{\rm max}} (l!)^m)$
where $S$ is the number of vertex permutations that transform the diagram into
itself, and $m$ is the number of equivalent $l$-tuples of lines. Equivalent
lines are lines that begin and end at the same vertices with the
same direction of arrows.

\end{enumerate}

We  illustrate the use of these Feynman rules by applying them
to some of the diagrams in Fig.~\ref{fig_hugenholtz}.
We start with the diagram at ${\cal O}(\kf^6)$, Fig.~\ref{fig_hugenholtz}(a).
This diagram has one vertex and one equivalent pair of lines
originating and ending at this vertex.
The overall factor is therefore $i/2$ and we have
\beq
 \edens_1 = \frac{i}{2}\,(-iC_0)\,g(g-1)\left(
    \lim_{\eta\to 0^+} \int\!\frac{d^4 k}{(2\pi)^4}\,e^{ik_0 \eta}\,iG_0(\kt)
    \right)^2 \ ,
\eeq
The $dk_0$ integration is performed using contour integration,
which picks up the $\theta(\kf-k)$ pole, and
the remaining $d^3 k$ integral up to $\kf$
is trivial.  The result is 
\beq
  \edens_1 =\dens\,(g-1)\,\frac{\kf^2}{2M}\,\frac{2}{3\pi}\, \kf a_s
         \ ,
\eeq
where Eqs.~(\ref{C2imatch}) and (\ref{fermid}) have been used \cite{Len29}. 
$\edens_1$ is of order $\kf^6$ as 
expected from the power counting. We label the energy contributions 
by the power of $\kf$ that accompanies the generic scale $R$;
this power is indicated by the subscript $j$ on $\edens_j$.

At ${\cal O}(\kf^7)$ we have two diagrams. Figure~\ref{fig_hugenholtz}(c)
vanishes at $T=0$ because it requires particle-hole pairs of equal
momentum at the Fermi surface, which do not exist.
(In particular, there are integrals of the form $\int\! dk f(k) \theta(\kf-k) 
\theta(k-\kf)$, which vanish if $f(k)$ is regular.)
Such diagrams are called ``anomalous.''
Figure~\ref{fig_hugenholtz}(b) has $S=2$ and two equivalent pairs of lines,
so the symmetry factor is $i/8$. Applying the Feynman rules we obtain
\beq
  \label{energy2a}
   \edens_2 = -i\,\frac{(C_0 )^2}{4}\,g(g-1)
    \int\!\frac{d^4 k}{(2\pi)^4}\int\!\frac{d^4 p}{(2\pi)^4}
    \int\!\frac{d^4 q}{(2\pi)^4}\,
	      G_0(\pt)\,G_0(\pt-\qt)\,G_0(\kt+\qt)\,G_0(\kt) \ .
\eeq
The frequency integrals are again simple contour integrals.
A transformation to dimensionless
center-of-mass variables 
$\vec{s}$, $\vec{t}$, and $\vec{u}$ using
\beq
   \vec{p}=\kf(\vec{s}+\vec{t}),\qquad \vec{k}=\kf(\vec{s}-\vec{t}), \qquad
      \mbox{ and }\qquad \vec{q}=\kf(\vec{t}-\vec{u})
\eeq
leads to 
\bea
  \edens_2 &=&\dens\,(g-1)\,\frac{\kf^2}{2M}\,(\kf a_s )^2 \frac{48}{(2\pi)^5}
       \int\! d^3 s \int\! d^3 t \int\! d^3 u \,\theta(1-|\vec{s}+\vec{t}|)
        \theta(1-|\vec{s}-\vec{t}|)   \nonumber \\
     & & \qquad\qquad \null \times
	      \theta(|\vec{s}+\vec{u}|-1) \theta(|\vec{s}-\vec{u}|-1)
            \frac{1}{t^2-u^2+i\epsilon}
     \ ,
   \label{energy2b}
\eea
where Eq.~(\ref{C2imatch}) has been used again.

The integral over $u$ in Eq.~(\ref{energy2b}) contains a linearly 
divergent term in $D=3$ dimensions
(putting back the factors of $\kf$):
\beq
       \int_{\kf}^\infty\! \frac{d^{D} u}{t^2-u^2+i\epsilon}=
       \int_0^\infty\! \frac{d^{D} u}{t^2-u^2+i\epsilon}-
       \int_0^{\kf}\! \frac{d^{D} u}{t^2-u^2+i\epsilon} \ .
       \label{eq:divergent}
\eeq
When dimensionally regularized, the first term on the right-hand
side is purely imaginary in our minimal subtraction prescription
[cf. Eq. (\ref{LnDR})]. Since the energy density is real, all imaginary
parts cancel in the end and the first term does not contribute.
We have verified this cancellation explicitly.
In the conventional approach, 
one avoids such power-law divergences by eliminating the potential, which 
is implicitly
assumed to be valid for arbitrarily high momenta, in favor of the 
scattering amplitude.
This amounts to summing the ``ladder'' diagrams 
to all orders and expanding the result afterwards.
The EFT deals with 
such power law divergences more directly by 
avoiding the assumption of an underlying potential. 
Power law divergences are simply 
subtracted out and the effect of high energy modes is captured in the 
coefficients of the terms in the effective potential.
Renormalization theory assures us that this procedure is consistent
and correct \cite{LEPAGE89}.
As in the conventional approach, one could
also resum all bubbles with $C_0$ contact interactions in the EFT
and apply this two-body scattering amplitude instead of the $C_0$ 
interaction at finite density. Since the
resummed amplitude falls like $1/p$ at large momenta, 
some ultraviolet divergences at finite density would be avoided. 
In the case of nuclear matter, where some mass scales are fine tuned
to be unnaturally large, such a resummation will have to be done.
(See Refs. \cite{KSW98,UvK99} for a discussion of this issue in 
the vacuum case.)
For natural scales, however, it is much easier to treat all interactions
perturbatively. As long as the renormalization is carried out consistenly,
the results are insensitive to the regularization scheme.

The second term in Eq.~(\ref{eq:divergent})
is finite and the remaining integrals 
can be evaluated analytically after some judicious  partial integrations
(see Appendix~C of Ref.~\cite{FST00} for details).
The net result for the second-order contribution to the energy density is
\beq
   \edens_2 =\dens \,(g-1)\,\frac{\kf^2}{2M}\,(\kf a_s)^2 \frac{4}{35\pi^2}
   \left(11-2\ln2 \right)   \, ,
                \label{energy2d}
\eeq
which was first obtained by Huang and Yang \cite{HuY57}.

At ${\cal O}(\kf^8)$, we have seven diagrams contributing to the energy
density. 
Their evaluation from the Feynman rules is straightforward
although somewhat tedious;
we will not go through the calculations in detail here but only 
make some comments. 
Three of the diagrams, 
Figs.~\ref{fig_hugenholtz}(d), \ref{fig_hugenholtz}(e), and 
\ref{fig_hugenholtz}(f),
vanish 
when evaluated. 
This is immediately clear for Fig.~\ref{fig_hugenholtz}(d),
which vanishes due to the same argument as the 
similar diagram at ${\cal O}(\kf^7)$. 
In Fig.~\ref{fig_hugenholtz}(e), the contributions
of hole and particle propagation in the line with the tadpole loop
attached cancel exactly. 
This cancellation is manifest once the 
frequency integrals have been carried out. 
Figure~\ref{fig_hugenholtz}(f) can again be seen to vanish
without any calculation. 
Assign a four-momentum $\kt$ to the center loop 
to which the three tadpole loops are attached and perform 
the $dk_0$ integration. We have to evaluate $\int\! dk_0\, G(\kt)^3$,
but only simple poles in $k_0$ contribute to the integral. The
terms with simple poles, however, involve a factor $\theta(k-\kf)
\theta(\kf-k)$ times a regular function and so the diagram vanishes.
Note that because the interactions are contact terms, a large 
number of exchange diagrams generally vanish as anomalous diagrams.

Diagrams \ref{fig_hugenholtz}(g) and (h), each with three
$C_0$ vertices, can be directly calculated.
The first one contains a linear divergence similar to the diagram at
${\cal O}(\kf^7)$, which can be isolated ({\it e.g.}, by transforming
to center-of-mass variables), and then 
removed by minimal subtraction, leaving a finite remainder.
The second diagram is finite. We have not obtained
analytic results  for these diagrams.  After the principal value
integrals are carried out analytically, however, the numerical
evaluation of the remaining integrals is straightforward.
We have calculated these integrals 
using a standard Monte Carlo integration routine \cite{vegas}
and reproduced the
values given in the literature \cite{AMUSIA68,BAKER71,BISHOP73}. 
Further details will be given elsewhere \cite{HAMMER00}.

Finally, there are diagrams  \ref{fig_hugenholtz}(i) and 
\ref{fig_hugenholtz}(j), 
which contain $C_2$ and $C_2'$ vertices.
These vertices contribute for the first time at ${\cal O}(\kf^8)$
and introduce a dependence on the $s$-wave effective range $r_s$ and 
the $p$-wave scattering length $a_p$.
The diagrams are very similar to the one at ${\cal O}(\kf^6)$
and can be evaluated directly.
One must be careful with the sign of the relative 
momenta $\vec{k}$ and $\vec{k}'$, since in the crossed contribution
$\vec{k}'$ changes its sign. This is irrelevant for the $C_2$
vertex, but for $C_2'$ it changes the overall factor $(g-1)$
into $(g+1)$.
Adding all contributions at ${\cal O}(\kf^8)$, we have
\bea
      \edens_3 &=&
	        \dens\,\frac{\kf^2}{2M}\Bigl[ (g-1)\frac{1}{10\pi}\,(\kf a_s)^2
        \,\kf r_s + (g+1)\,\frac{1}{5\pi}\,(\kf a_p)^3  
		      \nonumber \\
       & &  +(g-1)\{(0.07550\pm 0.00003)+(g-3)\,(0.05741\pm 0.00002)\}\,
           (\kf a_s)^3 \vphantom{\frac{1}{10\pi}} \Bigr] \,,
       \label{energy3}
\eea
which agrees with previous calculations \cite{AMUSIA68,BAKER71,BISHOP73}.

Equation~(\ref{energy3}) illustrates that a conventional argument
about the relative contribution of diagrams does 
not apply to the EFT-based diagrammatic expansion. 
Consider diagrams  \ref{fig_hugenholtz}(g) and  \ref{fig_hugenholtz}(h) again.
Diagram  \ref{fig_hugenholtz}(g) has terms with two and four hole
lines, while diagram \ref{fig_hugenholtz}(h) has only terms with three hole 
lines and should be relatively suppressed according to the argument 
\cite{holearg}. Equation~(\ref{energy3})
shows that both diagrams are of the same order of magnitude, as
expected from EFT power counting.
We return to this point in the discussion below.

In this section we have calculated the energy density of a dilute Fermi gas
up to order $(\kf R)^3$. In contrast to conventional calculations,
the EFT approach is completely transparent 
and easily extended to higher orders, finite temperatures, and other
observables.
The EFT method is particularly advantageous when logarithmic
divergences are present, as discussed in the next section.

\section{Logarithmic Singularities}    

A naive extrapolation of Eq.~(\ref{energy3}) might lead
one to conclude
that the energy density has an analytic expansion in $\kf/\Lambda$. 
This would be true if the Feynman graphs contained only power divergences. 
However, diagrams for three-to-three scattering
have logarithmic divergences, and
a consequence is that some terms in the energy
density appear multiplied by powers of $\ln\kf$.
Renormalization group methods provide a powerful tool to discern
this analytic structure. 
We follow the discussion of
Braaten and Nieto, who performed an analogous analysis for dilute
Bose gases \cite{BRAATEN97}. 

As we discussed in Sect.~\ref{FSEFT},
logarithmic divergences are special because they match onto logarithms
of physical quantities.
This means that we can identify where physical logarithms might appear
by examining
the renormalization group equations for the EFT coefficients.
The general structure of these equations for dimensional regularization
with minimal subtraction is
\beq
  \mu \frac{d}{d\mu}g_j(\mu) = \beta_j(g) \ ,
  \label{rge}
\eeq
where $\mu$ is the (arbitrary)
renormalization scale,
$g_j$ is a generic $n$-body coupling ({\it e.g.}, a $C_{2i}$ or a
$D_{2i}$), and $\beta_j(g)$ is a polynomial in the couplings $g_i$
with $\Lambda$-independent coefficients.
The terms in the polynomial correspond to diagrams which, if they
have logarithmic divergences, will have nonzero coefficients. 
Matching $\Lambda$ dimensions on each side of the equation
[recall Eq.~(\ref{scaling})] greatly restricts the possible 
terms \cite{BRAATEN95}. 

\begin{figure}[t]
\begin{center}
  \epsfxsize=12.cm
  \epsffile{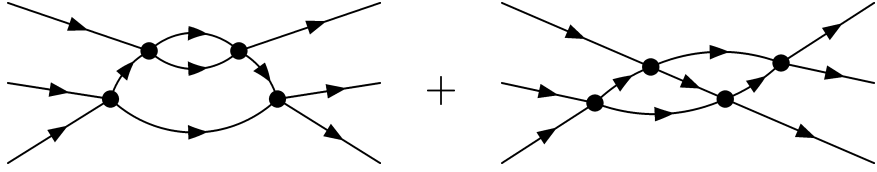}
\end{center}
\caption{Diagrams for three-particle scattering in the vacuum that
contribute to the running of $D_0$.}
\label{fig_3to3}
\end{figure} 

The running of two-body couplings $C_{2i}$ is determined
by graphs in the two-body sector where there are only power divergences;
therefore, none of these couplings depend on $\mu$.
So the lowest-dimension coupling that might run is $D_0$, the coefficient
of the three-body contact term 
in the EFT Lagrangian,
\beq
   {\cal L}_{\rm 3-body}=-\frac{1}{6}D_0(\psi^\dagger \psi)^3
          \ .
\eeq
Since $D_0 \sim 1/M\Lambda^4$ and  
$C_{2i}\sim 1/M\Lambda^{2i+1}$, $\beta_j$ could have four $C_0$'s or one
$C_0$ and one $C_2$ or $C_2'$.
The latter two correspond to finite tree diagrams, so their coefficients
in $\beta_j$ are zero.
That leaves only 2-loop diagrams with four $C_0$ vertices, and an
inspection of these uncovers logarithmic divergences in
the diagrams in Fig.~\ref{fig_3to3}.
 
Performing the loop integrals in $D<3$
dimensions and then analytically continuing back to $D=3$,
the logarithmic divergences leads to simple poles 
$1/(D-3)$ in the $3\to 3$ scattering amplitude.
Braaten and Nieto \cite{BRAATEN97} have extracted the pole part of
the diagrams in Fig.~\ref{fig_3to3} (which involves removing an overlapping
divergence). 
Using their
result and taking into account the diagrams with cyclic 
permutations of the external momenta, the pole part
of the $3\to 3$ scattering amplitude is
\bea
i[{\cal T}_{3\to 3}]_{\rm pole}
   &=& -iM^3 \,(C_0 )^4 \,\frac{4\pi-3\sqrt{3}}{8\pi^3}
      \,\frac{1}{D-3}\,\mu^{2(3-D)}
	  \nonumber \\
	&=&
	  -iM^3 \,(C_0 )^4 \,\frac{4\pi-3\sqrt{3}}{8\pi^3}
	  \biggl[ \frac{1}{D-3} - 2\ln\mu + \cdots \biggr] \ ,
    \label{3to3coeff}
\eea
where the spin structure has been  suppressed and
the omitted terms vanish as $D\rightarrow 3$.
A counterterm removes the $1/(D-3)$ pole   in Eq.~(\ref{3to3coeff}),
leaving a finite term $[{\cal T}_{3\to 3}]_{\rm finite}$,
which  depends on the arbitrary scale $\mu$.
Since physical quantities cannot depend on 
$\mu$, the renormalized coupling $D_0$ must satisfy
\beq
     \frac{d}{d\mu}\Bigl(
            i [{\cal T}_{3\to 3}]_{\rm finite}
		   - i D_0 \Bigr)=0 \ ,
			\label{rgeD}
\eeq
or
\beq
\mu\frac{d}{d\mu}D_0=M^3\, (C_0 )^4\, \frac{4\pi-3\sqrt{3}}{4\pi^3}\,,
\eeq
where the limit $D\to 3$ has been taken.%
\footnote{Note that there are other diagrams contributing
to $[{\cal T}_{3\to 3}]_{\rm finite}$.  These diagrams, however, have no
logarithmic divergences and consequently no $\mu$ dependence.}
This has the form of Eq.~(\ref{rge}).

Equation~(\ref{rgeD}) is trivially integrated to find $D_0(\mu)$:
\beq
\label{d0run}
D_0(\mu)=D_0(1/a_s)+M^3\, (C_0 )^4 \,\frac{4\pi-3\sqrt{3}}{4\pi^3}\,
\ln(a_s \mu)\,,
\eeq
where we have chosen our reference scale to be the natural $s$-wave
scattering length $a_s$. 
Changes in the renormalization scale $\mu$ correspond to
changes in how well the scattering at short distances is resolved. 
As less of the successive two-body scatterings in Fig.~(\ref{fig_3to3})
is resolved, the contributions are shifted to the three-body
contact term $D_0$.

We can now use Eq.~(\ref{d0run}) to obtain the contribution
of order $\kf^9 \ln\kf$ to the energy density of the dilute Fermi
gas. In order to obtain the full
contribution at ${\cal O}(\kf^9)$, we would have to renormalize the 
$3\to 3$ amplitude in the vacuum, perform the matching 
to determine $D_0(1/a_s)$
and then 
calculate all the energy diagrams at this order.
(Alternatively, we could perform the matching with data from a
many-body system.)
To obtain just the nonanalytic contribution at this order, however,
we have only to calculate diagram $(k)$
in Fig.~\ref{fig_hugenholtz}, since the logarithm in
Eq.~(\ref{d0run}) must match to the $\ln\kf$ from diagram $(l)$ (plus
others) \cite{BRAATEN95}.
Equivalently, we keep only Fig.~\ref{fig_hugenholtz}$(k)$ with $D_0$
evaluated at 
$\mu=\kf$, which avoids large cancellations between the diagrams.
Applying the Feynman rules from the previous section we then have
\beq
\edens_4^{\ln}=\dens\,(g-2)(g-1)\,\frac{\kf^2}{2M}\,\frac{16}{27\pi^3}
\,\left(4\pi-3\sqrt{3}\right)\,(\kf a_s)^4 \, \ln(\kf a_s)\,,
\eeq
which is the result obtained previously \cite{AMUSIA68,BAKER71,BISHOP73}.
Note that more than a two-fold degeneracy ($g>2$) 
is needed to get a non-zero contribution, which follows
from  Pauli principle restrictions.
This procedure can be continued to higher orders to identify 
further $(\ln\kf)^n$ terms \cite{BRAATEN97}.

\section{Discussion and Summary}

We have derived the energy per particle in a dilute Fermi gas
through terms of order $\kf^6 \ln \kf$, reproducing results 
given in Refs.~\cite{AMUSIA68,BAKER71,BISHOP73}.
This calculation illustrates many of the virtues of an EFT
approach to many-body physics.
The EFT provides a model independent description of finite density
observables in terms of parameters that can be fixed from
scattering in the vacuum, with no off-shell ambiguities.
Since the system is probed at low resolution,
all vertices are contact interactions, which means
exchange contributions are trivial and many diagrams vanish.
After taking advantage of our freedom to choose a subtraction scheme,
the power counting is exceptionally clean:  each diagram contributes
to a single, well-defined order in the low-density expansion.
Finally, the EFT provides tools for identifying and understanding
the analytic structure of observables
in terms of the short- (and long-) distance behavior.
Future extensions of this work will include the generalization to finite 
temperature, single-particle properties, and linear response, as well as 
the complete renormalization at ${\cal O}(\kf^9)$ \cite{HAMMER00}.

One can compare the clean, minimal calculation using an EFT to
previous calculations, which required a more general set of diagrams
that needed to be summed and expanded, with care taken to avoid
double counting. 
In the standard treatments \cite{AMUSIA68,BAKER71,BISHOP73},
the focus is on the potential, rather than
on observables as in the EFT.
The EFT perspective says directly that
at low energy the effective range parameters matter and 
not the details of a potential.
A corollary is that one can use any complete set of regularized local
interactions that
reproduces the low-energy observables; there is no ``best'' potential.
It's not that interactions {\it are\/} pointlike at low resolution,
but that true observables are {\it indistinguishable\/}
(up to a well-defined truncation error) from those calculated
with pointlike interactions.
These interactions are wrong at short distances, but renormalization
theory prescribes how the discrepancy from incorrect high energy behavior
can be systematically corrected by the renormalization of the
EFT coefficients.

The idea of replacing the ``true'' potential with one that 
sums low-energy scattering effects and is easier
to work with (a ``pseudo-potential'') is an old idea \cite{pseudopot}.
The EFT framework makes it systematic and extendable.
Even if we {\it did\/} know the underlying interaction ({\it e.g.},
a hard sphere potential), it is simpler and more efficient to match in 
the vacuum to an EFT and then to use the EFT for the finite density 
calculation. This is analogous to the use of effective theories for
heavy-quark physics (NRQCD) and QED bound states (NRQED), which are 
beneficial even though the underlying theories are known (cf. 
Ref.~\cite{LEPAGE89}).

The EFT analysis can also clarify the nature of many-body systems.
In Ref.~\cite{FETTER71}, the  energy expansion for a dilute Fermi gas
is followed by a discussion of the hole-line expansion, which is used
to do a form of power counting.
The simple argument is that each additional hole-line is accompanied
by momenta restricted by $\kf$, leading to
an additional factor of density that suppresses the diagram at
low density. Particle lines have unrestricted momenta and are 
therefore not suppressed, and so must be included to all orders.
The power counting in our EFT reveals that a hole-line
expansion for the low-density natural Fermi gas is only an artifact
of a subtraction scheme that does not remove power divergences.
Indeed, the particle-particle and particle-hole
ring diagrams in Fig.~\ref{fig_hugenholtz} contribute to the same order
in $\kf$ and are numerically comparable.
This is cleanly reproduced in the EFT because of minimal subtraction,
which results in no distinction between particle and hole lines 
in the power counting.

We saw that going
further in the dilute Fermi gas
energy expansion requires input from three-to-three scattering
(or other many-body observables).
This means that three-body (and four-body) contributions are
{\em inevitable\/}.
If only two-body potentials were used, one would find different results
at this order in the expansion for different potentials that agreed for
two-body scattering, which would conventionally be attributed
to a lack of knowledge of the off-shell two-body potential.
From the EFT perspective, the differences would be due to short-distance
behavior that must be corrected with many-body counterterms. 
For the three-nucleon system, accounting for the three-body contact
interaction leads to a compelling explanation of the 
Phillips line \cite{BEDAQUE98}.
The need to renormalize because of incorrect high-energy behavior
is not avoided because of finite loop integrals ({\it e.g.}, due to
a form factor). A sensitivity to cutoff parameters is a signature that 
something is missing in the theory.

The extension of the EFT procedure for dilute, repulsive Fermi systems
to the nuclear matter problem \cite{lutz,STEELE99} is far from immediate.
In particular,
the  EFT subtraction and renormalization scheme used here is not appropriate
for a treatment of nuclear matter.
That is because the EFT momentum
breakdown scale will be set by the lowest mass scale in the effective
range expansion, which is of order 10\,MeV because of the fine-tuned
scattering lengths. Since nuclear matter equilibrium density is around 
$\kf \approx 270\,$MeV, the expansion parameter $\kf/\Lambda\approx 27$
and the perturbation series is not useful.
Alternative power counting schemes have been proposed to deal with this 
problem in the vacuum \cite{KSW98,UvK99}.
The consistent extension of such a scheme to finite density, which may
reintroduce a hole-line expansion, is the next step. However,
an EFT with only short-range interactions will still fall far short, 
with a breakdown scale of $m_\pi/2$ \cite{STEELE99}.
Thus, pions will have to be included as long-range physics.
We are confident 
that the virtues of the EFT methods illustrated for a much simpler
system here will continue to pay dividends as they are
adapted to the study of nuclear matter.

\vskip 0.6cm
{\bf \noindent  Acknowledgements \hfil}
\vglue 0.4cm
We acknowledge useful discussions with E.~Braaten, J.~Ho, G.~P.\ Lepage,
  R.~J.\ Perry, B.~D.\ Serot, J.~V.\ Steele and N.~Tirfessa.
This work was supported by the National Science Foundation
under Grant No.\ PHY--9800964. HWH thanks the Institut
f{\"u}r Kernphysik at the University of Mainz for its hospitality 
during completion of this work.

\end{document}